# Giant impacts in the Saturnian System: a possible origin of diversity in the inner mid-sized satellites

**Yasuhito Sekine**

Dept. Complexity Sci. & Engr., University of Tokyo

5-1-5 Kashiwanoha, Kashiwa, Chiba 277-8561, Japan

Email: sekine@k.u-tokyo.ac.jp

Tel & Fax: +81-4-7136-3954

**Hidenori Genda**

Dept. Earth & Planet. Sci., University of Tokyo

7-3-1 Hongo, Bunkyo, Tokyo 113-0033, Japan

Email: genda@eps.s.u-tokyo.ac.jp

**Abstract**: It is widely accepted that Titan and the mid-sized regular satellites around Saturn were formed in the circum-Saturn disk. Thus, if these mid-sized satellites were simply accreted by collisions of similar ice-rock satellitesimals in the disk, the observed wide diversity in density (i.e., the rock fraction) of the Saturnian mid-sized satellites is enigmatic. A recent circumplanetary disk model suggests satellite growth in an actively supplied circumplanetary disk, in which Titan-sized satellites migrate inward by interaction with the gas and are eventually lost to the gas planet. Here we report numerical simulations of giant impacts between Titan-sized migrating satellites and smaller satellites in the inner region of the Saturnian disk. Our results suggest that in a giant impact with impact velocity ≥ ~1.4 times the escape velocity and impact angle of ~45 degree, a smaller satellite is destroyed, forming multiple mid-sized satellites with a very wide diversity in satellite density (the rock fraction = 0–92 wt%). Our results of the relationship between the mass and rock fraction of the satellites resulting from giant impacts reproduce the observations of the Saturnian mid-sized satellites. Giant impacts also lead to internal melting of the formed mid-sized satellites, which would initiate strong tidal dissipation and geological activity, such as those observed on Enceladus today and Tethys in the past. Our findings also imply that giant impacts might have affected the fundamental physical property of the Saturnian mid-sized satellites as well as those of the terrestrial planets in the solar system and beyond.

**Key words:** Icy satellites; Giant impact; Satellite formation; Differentiation; Saturnian system

# 1. Introduction

The regular satellites around Jupiter, Saturn, and Uranus are often viewed as a miniature solar system (e.g., Lunine and Stevenson, 1982; Stevenson et al., 1986; Mousis et al., 2002; Mosqueira and Estrada, 2003a; Canup and Ward, 2006; Lunine et al., 2009). These satellites orbit approximately within the planet's orbital plane with very low eccentricities and inclinations and are considered as a by-product of the gas planet's formation. In the standard picture of gas planet formation, the last stage of the formation of the gas planets involves an inflow of gas and solids from the solar nebula and the formation of circumplanetary disks (e.g., Pollack et al., 1996; Canup and Ward, 2006). In the circumplanetary disk, regular satellites would have been accreted by collisions of rock-ice satellitesimals (e.g., Lunine and Stevenson, 1982). Jupiter and Uranus each have four similarly-sized large satellites (Fig. 1). It is suggested that the systematic progression of density in the Galilean satellites (Fig. 1) reflects higher rock fraction in the satellitesimals due to higher temperatures in the inner Jovian subnebula (e.g., Lunine and Stevenson, 1982).

Unlike the Jovian and Uranian systems, the Saturnian system seems to be enigmatic from the viewpoint of a miniature solar system. This is because (1) Saturn has the only one large satellite (i.e., Titan), which accounts for ~97% of the total mass of the Saturnian system, and (2) the mid-sized satellites (i.e., satellites with radii between 100 and 1000 km) have a wide diversity of density independently of their orbital distances from Saturn (Fig. 1) (Matson et al., 2009). In fact, Tethys has a surprisingly low density close to that of pure ice (0.973 g/cm$^3$; rock fraction = 6 wt%), whereas the densities of Enceladus (1.608 g/cm$^3$; rock fraction = 57 wt%) and Dione (1.476 g/cm$^3$; rock fraction = 50 wt%) suggest their rock-rich compositions (Jaumann et al., 2009; Matson et al., 2009). These aspects of the Saturnian system may be difficult to be explained by satellite formation from rock-ice satellitesimals in a simple minimum mass disk, implying an importance of dynamic and stochastic events in the formation of the Saturnian system.

Recently, the gas-starved model (e.g., Canup and Ward, 2002; 2006; Alibert et al., 2005; Alibert and Mousis, 2007; Sasaki et al., 2010) proposes satellite growth in an actively supplied circumplanetary disk, sustained by an inflow from the solar nebula. Large satellites formed in the inflowing gas disk migrate inward by interaction with the gas (the type I migration) and are eventually lost to the gas planet (Alibert et al., 2005; Canup and Ward, 2006; Sasaki et al., 2010). The gas-starved model suggests that such growth and loss of large satellites continuously occur in the disk until the gas flow ends (Canup and Ward, 2006). According to the gas-starved model, a Saturn-like satellite system can be produced when large inner satellites are lost by spiraling into the planet and another large outer satellite (i.e., Titan) remains in the disk as the gas inflow ends (Canup and Ward, 2006).

Canup (2010) recently indicates that when a differentiated large satellite migrates into the Roche limit of Saturn, the icy mantle of the satellite would have been removed by Saturn's tidal force, resulting in the formation of Saturn's pure-ice ring and ice-rich Tethys. These scenarios can account for why Titan is the only large satellite in the Saturnian system and why Saturn's ring and Tethys are extremely enriched in ice components. However, the mechanism responsible for the diversity of density (especially, high density) in the mid-sized satellites remains mystery. Furthermore, if the above scenario is the case, Tethys would have been formed from massive icy materials spreading right outside the Roche limit and then evolved outward because of tidal interaction with Saturn (Canup, 2010). This picture might be inconsistent with the presence of primitive Mimas and rock-rich Enceladus orbiting inside Tethys.

Here, we propose an alternative mechanism to explain the wide diversity of density in the inner mid-sized satellites, including ice-rich Tethys and rock-rich Enceladus and Dione, based on the gas-starved disk model. When large satellites migrate in the disk toward Saturn, "giant impacts" between a migrating satellite and an inner satellite may have occurred. Such a dynamic process might have formed a rock-rich satellite and/or an ice-rich satellite by blowing off ice mantle of the satellites, as suggested for the formation of metal-rich Mercury and rock-rich Moon in the inner solar system (Benz et al., 1987; 1988; Cameron, 1997; Canup and Asphaug, 2001). Fragmentation of a differentiated satellite also might lead to the formation of small satellites with a wide diversity in density. After the migrating satellite was lost to Saturn, small satellites with various rock fractions would have been left in the inner region of the disk. In this study, we investigate whether such giant impacts are capable of producing the diversity in satellite density using a three-dimensional smooth particle hydrodynamic (SPH) simulation. We also discuss the impact conditions and viability of such giant impacts by comparing the numerical results of post-impact small satellites with the Saturnian mid-sized satellites.

## 2. Numerical model

In order to investigate giant impacts between icy satellites, we use the SPH method (e.g., Monaghan, 1992). In the SPH method, objects are described by a multitude of overlapped particles called SPH particles. The physical property of each SPH particle is tracked in time in a Lagrangian manner. SPH particles are evolved due to compressive heating, expansive cooling, shock energy dissipation, and gravity. Because the SPH method can treat large-scale deformation and shock wave propagation, it has been widely used for hydrodynamic simulations of hypervelocity impacts, such as cometary impacts on icy satellites (e.g., Bruesch and Asphaug, 2004; Fukuzaki et al., 2010; Sekine et al., submitted), impacts of asteroids (e.g., Durda et al., 2004; Jutzi et al., 2008), and giant impacts of terrestrial planets and dwarf planets (e.g., Benz et al., 1987; Cameron and Benz, 1991; Canup and Asphaug, 2001; Canup, 2005; Asphaug et al., 2006).

Considering large uncertainties in the impact conditions of the giant impacts (i.e., the impact velocity ($v_{imp}$), impact angle ($\theta_{imp}$), rock-to-ice mass ratio in the pre-impact satellites, and impactor-to-target mass ratio), we systematically performed 20 simulations of giant impacts for various parameters. We changed $v_{imp}$ and $\theta_{imp}$ from 1.4 to 2.6 times the escape velocity ($v_{esc}$) and from 30 to 60 degrees, respectively. The impact with $\theta_{imp}$ of 0 degree corresponds to a head-on collision.

The rock-to-ice ratio of icy satellitesimals varies depending on the disk temperature and chemical composition of C- and N-bearing volatiles (e.g., Prinn and Fegley, 1989). Recent theoretical and experimental studies suggest that $CO_2$ and $NH_3$ are major C- and N-bearing volatiles in the circumplanetary disk, respectively (e.g., Sekine et al., 2005; Alibert and Mousis, 2007). We consider that all of $CO_2$, $CH_4$, and $NH_3$ are incorporated in ice components forming pure ices, clathrates, and hydrates at temperatures of the Saturnian subnebula (e.g., 50–90 K) (e.g., Sekine et al., 2005; Alibert and Mousis, 2007). When using $CO_2:CO:CH_4$ = 30:10:1 and $NH_3:N_2$ = 1:1 in the relative abundances (Alibert and Mousis, 2007) and the solar abundance (Anders and Grevesse, 1998), the rock-to-ice ratio in icy satellitesimals formed at 50–90 K is calculated approximately 3:7. In the present study, we assume that the rock-to-ice mass ratio in both the impactor and target is same as that of icy satellitesimals (i.e., rock : ice = 3:7).

The rock fraction of the target may have been larger than 30 wt%, given a massive escape of $H_2O$ during accretion (Kuramoto and Matsui, 1994). In the present study, however, the impactor size was assumed to be much smaller than that of the target (see below). Accordingly, such a difference in the initial rock fraction of the target satellite does not change our conclusion significantly. Based on the gas-starved model (Canup and Ward, 2006), we conducted numerical simulations for a Titan-sized satellite as the migrating target satellite. In contrast to the target satellite, the mass of the impactor satellite is uncertain. Given the current mass of the mid-sized satellites, we assumed the impactor-to-target mass ratio of 1/20. As a first approximation, we also assumed that both the target and impactor were differentiated. The effects of the satellite size and differentiation on the calculation results are discussed below in Sec. 4.1.

The mutual gravitational forces among all SPH particles are directly computed using the GRAPE-6A system (Fukushige et al., 2005), which is specialized for gravitational N-body calculations. Each SPH simulation is performed with $2 \times 10^4$ and $5 \times 10^4$ SPH particles for 17 and 3 runs, respectively. In the present study, the Tillotson equation of state (Tillotson, 1962) is used with material constants of granite for rock material and water ice for ice material. Material strength is not considered in our simulations, because a fluid description seems to be valid in giant impacts given that stresses associated with self-gravity exceed the material strength (Canup, 2006).

## 3. Results

Our numerical simulations show that there are three regimes of giant impacts between a Titan-sized satellite and a mid-sized satellite; merging collisions, hit-and-run collisions, and clump-forming collisions. A merging collision means that an impactor simply merges to a target satellite. In a hit-and-run collision, an impactor does not merge with a target satellite and escapes from the target satellite after the impact, as proposed by the previous study of giant impacts between protoplanets (Agnor and Asphaug, 2004). A hit-and-run collision causes a significant change in the rock fraction of the impactor, because the outer icy layer of the impactor is usually stripped by the impacts (Asphaug et al., 2006). In a clump-forming collision, an impactor is largely destroyed and is divided into multiple "clumps". Here we define a "clump" as an aggregate of more than 10 SPH particles, which corresponds to the size of the mid-sized satellites with radii > 100 km. SPH particles belonging to a clump are gravitationally bounded with each other. However, multiple clumps formed by an impact are not gravitationally bounded and escape each other. Although the number of satellites does not change in hit-and-run collisions, clump-forming collisions result in increases in the number of satellite.

Figure 2 shows a typical numerical result of time series snapshots for a clump-forming collision. A clump-forming collision typically produces two to five mid-sized satellites with a wide diversity in the rock fraction. In fact, Fig. 2 shows that the impact produces two rock-rich clumps (clumps 1 and 2; the rock fraction = 44 and 48 wt%, respectively)) and two pure ice clumps (clump 3 and 4; the rock fraction = 0 wt%). Our numerical simulations show that impactor's icy materials on the side of impact point and rocky materials are ejected in relatively low velocities and then produce the rock-rich clumps. Meanwhile, icy materials on the opposite side of an impactor are less shocked and retain high downrange velocities, forming the ice-rich clumps. Although the outer icy layer of the target satellite near the impact point is also blown off by the clump-forming collision, the major constituents of the clumps are originated from the impactor.

Our numerical results show that merging collisions occur in impacts at low $v_{imp}$ ($v_{imp}$ < ~1.4 $v_{esc}$: Canup, 2005) and/or at low $\theta_{imp}$ (i.e., near head-on impacts (≤ ~30 degree)), whereas hit-and-run and clump-forming collisions take place in impacts at relatively high $v_{imp}$ (i.e., $v_{imp}$ ≥ ~1.4 $v_{esc}$). A key parameter for determining whether clump-forming collision or hit-and-run collision occurs is $\theta_{imp}$. In impacts at high $\theta_{imp}$ (≥ ~60 degree), impactors generally survive after giant impacts, resulting in hit-and-run collisions. Clump-forming collisions tend to occur in impacts at moderate $\theta_{imp}$ (~45 degree). This is probably because the internal energy of an impactor reaches the specific energy required for the fragmentation in an impact at moderate $\theta_{imp}$. On the other hand, impacts at high $\theta_{imp}$ subject impactors to much lower shock heating (i.e., much lower internal energy) than impacts at moderate and high $\theta_{imp}$. Given the random velocities expected in giant impacts in the case for terrestrial planet formation (~1–3 $v_{esc}$) (Agnor et al., 1999) and the distribution of $\theta_{imp}$ (sin 2$\theta_{imp}$ with 45 degree being the most likely value), our results would suggest that clump-forming collisions are highly viable in giant impacts between Titan-sized and mid-sized satellites.

Figure 3 shows our numerical results for the 20 runs of the relationship between the mass and rock fraction of post-impact satellites compared with those of the inner mid-sized satellites of the Saturnian system (Mimas, Enceladus, Tethys, Dione, and Rhea). In hit-and-run collisions, impactors lose ~15–50% of icy materials originally contained in their outer icy layer, because the outer icy layer is stripped by the impacts. Thus, hit-and-run collisions tend to increase the rock fraction of impactors and decrease the satellite mass in Fig. 3. In contrast, clump-forming collisions produce small satellites (the mass ranging from ~$10^{19}$ to $10^{21}$ kg) with a very wide range of the rock fraction (the rock fraction = 0–92 wt%). In comparison with the mass and rock fraction of the Saturnian mid-sized satellites, our results suggest that the giant impacts reproduce the observed wide diversity in the density of the Saturnian mid-sized satellites.

## 4. Discussion

### 4.1. Effects of the satellite size and differentiation

In the present study, we fixed the impactor-to-target mass ratio as 1/20 in the numerical simulations of the giant impacts. The impactor mass would be an essential parameter for determining whether a clump-forming collision occurs or not, because specific energy for fragmentation of the impactor would increase with the impactor mass in the gravity regime due to self-compression (Housen and Holsapple, 1990). This means that higher $v_{imp}$ would be required for larger impactors to cause clump-forming collisions. The previous study on collisional fragmentation proposes that there is a power law relationship between the specific energy at the fragmentation threshold ($Q^*$) and the radius of body ($R$) in the gravity regime ($Q^*$ is proportional to $R^{1.65}$) (Housen and Holsapple, 1990). Our calculations show that clump-forming collisions occur in impacts at $v_{imp} \geq 1.4\ v_{esc}$ and $\theta_{imp} = 45$ degree (Fig. 2). When using the power law relationship given by Housen and Holsapple (1990) as a first approximation, $v_{imp}$ required for a clump-forming collisions is higher than 2.2 $v_{esc}$ at $\theta_{imp} = 45$ degree for impactors with 25% of Titan mass. Considering the random velocity expected (~1–3 $v_{esc}$), clump-forming collisions would occur for impactors with ≤ 70% of Titan mass.

An important uncertainty in initial conditions is that we assumed fully differentiated impactors. Our simulations show that stripped outer layer and inner core of an impactor produce different satellites in a clump-forming collision (Fig. 2). Thus, even if an impactor is partially differentiated (i.e., only the outer layer is differentiated), satellites resulting from giant impacts would have a diversity in their densities. Given undifferentiated inner core of the partially differentiated impactor, highly rock-rich satellites (e.g., the rock mass fraction > 80 wt%) might not be produced. It is difficult to predict the degree of differentiation for pre-impact mid-sized satellites during accretion. The accretion of the inner mid-sized satellites is considered to be a rapid process (<$10^5$ years) (Mosqueira and Estrada, 2003a; Canup and Ward, 2006). According to the accretion model of icy satellites (Kuramoto and Matsui, 1996), the surface temperature of a growing satellite with 5% of Titan mass is calculated to be around 200 K. This temperature is too low to melt water ice but higher than the water-ammonia eutectic melting point of 176 K. Thus, if the ammonia concentration is significant in the satellites, the outer layer of impactor-sized satellite would be partially differentiated. Furthermore, when considering an impactor with 25% of Titan mass, the surface temperature would reach ~300 K, leading to a substantial melting of the outer layer (Kuramoto and Matsui, 1996). We accordingly consider that if impactors are larger than several percent of Titan mass, diversity in density of satellites resulting from giant impacts would be produced.

Based on the above discussions, we conclude that larger impactors are better for producing a wide diversity in density of post-impact satellites, because they are highly likely partially differentiated; nevertheless, larger impactor satellites require higher $v_{imp}$ to cause clump-forming collisions. To investigate such a trade-off relationship and viability to reproduce diversity in satellite density more quantitatively, further impact simulations varying the impactor size and degree of differentiation are needed.

## 4.2. Implications for evolution and exploration of the mid-sized satellites

Our simulations demonstrate that a clump-forming collision produces multiple satellites with a wide diversity in the rock fraction (Fig. 3); however, all of these satellites may not survive in long-term orbital evolution. Giant impacts would change the eccentricities and inclinations of the post-impact satellites, affecting their orbital evolutions (Kokubo and Genda, 2010). Planetary and satellite systems with multiple bodies also require a wide radial separations (> 3.5 times the Hill radius of the satellite) for stability (Lissauer, 1999; Canup and Ward, 2006). Thus, some of these satellites would have been lost by future collisions. Survival of the post-impact satellites strongly depends on their obtaining orbits that do not cross that of Titan-sized migrating satellite. To investigate long-term orbital evolution and survival of post-impact satellites, N-body simulations in the Saturnian disk coupled with SPH impact simulations are required; nevertheless, this is beyond the scope of the present study.

On the basis of the gas-starved model (e.g., Canup and Ward, 2002, 2006; Alibert and Mousis, 2007), we discuss giant impacts triggered by the migration of large satellites. However, the migration of large solid bodies due to gas drags and tidal torques commonly would occur in a gaseous disk (e.g., Ward, 1997). While large scale migration of Titan-sized satellites is explicitly advocated in the gas-starved disk model (Canup and Ward, 2002, 2006; Alibert and Mousis, 2007), it is possible that migration of large satellites might also occur in other satellite models as well (e.g., Mosqueira and Estrada, 2003b). Nevertheless, such migration has not been quantitatively modeled in the context of these other models.

Giant impacts are a potentially important process for the thermal evolution of the mid-sized satellites. If a giant impact has occurred on a mid-sized satellite, it would be the predominant heat source in the early stage of the satellite compared with other heat sources, such as accretion and decay of short-lived radioactive isotopes (Matson et al., 2009). In fact, our impact simulations show that the interior temperature of post-impact satellites becomes high sufficient to lead to internal melting and formation of deep oceans. Because presence of interior oceans in icy satellites causes strong tidal dissipation and heating in the oceans (Tyler, 2008), the impact-induced melting of the post-impact satellites would, in turn, trigger a long-lived self-maintaining warm interior and high levels of geological activities (Spencer et al., 2009). In addition, the high rock fractions in the satellites resulting from giant impacts are preferred for effective heating by the decay of radioactive isotopes, such as $^{26}$Al and $^{60}$Fe (Schubert et al., 2007). Accordingly, geological and geophysical observations of the mid-sized satellites may provide clues whether giant impacts have occurred in the Saturnian system.

In fact, the Cassini spacecraft is starting to reveal that the Saturnian mid-sized satellites have diversity not only in the rock fraction but also in the internal structure and levels of activity (e.g., Matson et al., 2009; Jaumann et al., 2009). Based on the gravitational data, Rhea is considered to be an undifferentiated body (Anderson and Schubert, 2010). Mimas does not show any clear evidence for recent or past geological activity, demonstrating a cold, undifferentiated interior (Matson et al., 2009; Jaumann et al., 2009). These satellites have the rock fractions explained by the solar composition (i.e., the rock mass fraction = ~30 wt%). On the other hand, Dione and Tethys have undergone endogenic activity, such as cryovolcanisms (Matson et al., 2009; Jaumann et al., 2009). In order to form a large rift that runs over 2000 km in length (Ithaca Chasma) on Tethys, it is suggested that a significant heat flux (18–30 mW/m$^2$) is required (Giese et al., 2007). The proposed heat flux is surprisingly high for Tethys, given its small size and very small rock fraction. Although such a high heat flux could have been achieved by tidal heating during Tethys' passage through an orbital resonance with Dione in the long-period orbital evolution (Chen and Nimmo, 2008), tidal dissipation associated with orbital evolution is a very complex process, which is strongly dependent on the initial condition. Enceladus is one of the most active satellite in the solar system (Porco et al., 2006), indicative of that its interior is probably differentiated (Schubert et al., 2007). Although the current warm and dissipation state is probably maintained by tidal heating with Dione, it is unclear how Enceladus first entered to its self-maintaining warm interior (Spencer et al., 2009). It is argued that initial heating from non-tidal heat sources is probably required for Enceladus to enter the current self-maintaining warm state (Spencer et al., 2009). Given these observations by the Cassini spacecraft, it seems that the satellites with an unusual rock fraction (i.e., Enceladus, Tethys, and Dione) show signs of high levels of activity today or in the past. Giant impacts could have been responsible for the heat sources triggering their activities. Mimas and Rhea might have survived in the migration of a Titan-sized satellite, retaining the original rock-ice ratio of the satellitesimals.

Given that the Cassini mission is expected to be continued until 2017, there will be more data becoming available on the properties of the Saturnian mid-sized satellites. Until now, Rhea is the only mid-sized satellite for which we have data on the quadrupole gravitational field. To better understand the early history of the Saturnian system, it is important to obtain gravitational data for the other mid-sized satellites. To assess whether Enceladus, Tethys, and/or Dione have been undergone giant impacts, it may be essential to compare chemical compositions of volatile species among the mid-sized satellites. In a giant impact, highly volatile species would be degassed and eventually escape from the satellites. Thus, giant impacts would have caused a contrast in the chemical compositions among the mid-sized satellites. It is suggested that intact crustal materials may be found on the wall of fresh and bright impact craters (Hibbitts et al., 2002). Thus, detailed observations of volatiles on such craters in the Cassini extended mission and/or future missions would provide an insight into the early history of the Saturnian system.

## 5. Summary

Our SPH simulations demonstrate that a giant impact between a Titan-sized satellite and a smaller differentiated satellites produce single or multiple mid-sized satellites with a wide diversity in the rock fraction. Physical properties of the post-impact satellites (such as the rock fraction) in the simulations reproduce those of the Saturnian mid-sized satellites. The recent Cassini's observations have revealed activities on Enceladus today and Tethys in the past. Both are conceivably linked to tidal heating, but the initial heat sources might have been needed to cause strong tidal dissipation and heating. Because giant impacts could have resulted in internal melting, they could have been the initial heat sources and may account for the linkage between the unusual rock fraction and presence of geological activities on the mid-sized satellites.

According to the planetary formation models, it is suggested the final stage of the formation of terrestrial planets and super-Earths around solar-type stars involves multiple giant impacts between protoplanets (e.g., Wetherill, 1985; Kokubo and Ida, 1998, 2007; Ida and Lin, 2010; Kokubo and Genda, 2010). Such giant impacts also would have played a key role for determining the iron-rock ratios (e.g., Marcus et al., 2010a) and water contents (e.g., Genda and Abe, 2005; Marcus et al., 2010b) of both terrestrial planets and super-Earths. Although it is suggested that Mercury and Moon have been formed by giant impacts (e.g., Benz et al., 1988; Cameron, 1997), the physical and chemical processes of giant impacts remain poorly constrained by geological and geophysical evidence. If giant impacts also have occurred on some of the Saturnian mid-sized satellites, detailed observations in future exploration of the mid-sized satellites would provide clues to an understanding of the nature of giant impacts, which might help to predict the physical and chemical properties of terrestrial planets and super-Earths beyond the solar system.

## Acknowledgements
Y.S. thanks S.A. Haider for the arrangement of the special issue; M. Higuchi for help in figure graphics. Support from Grants in Aid from Japan Society for the Promotion of Science is also acknowledged.

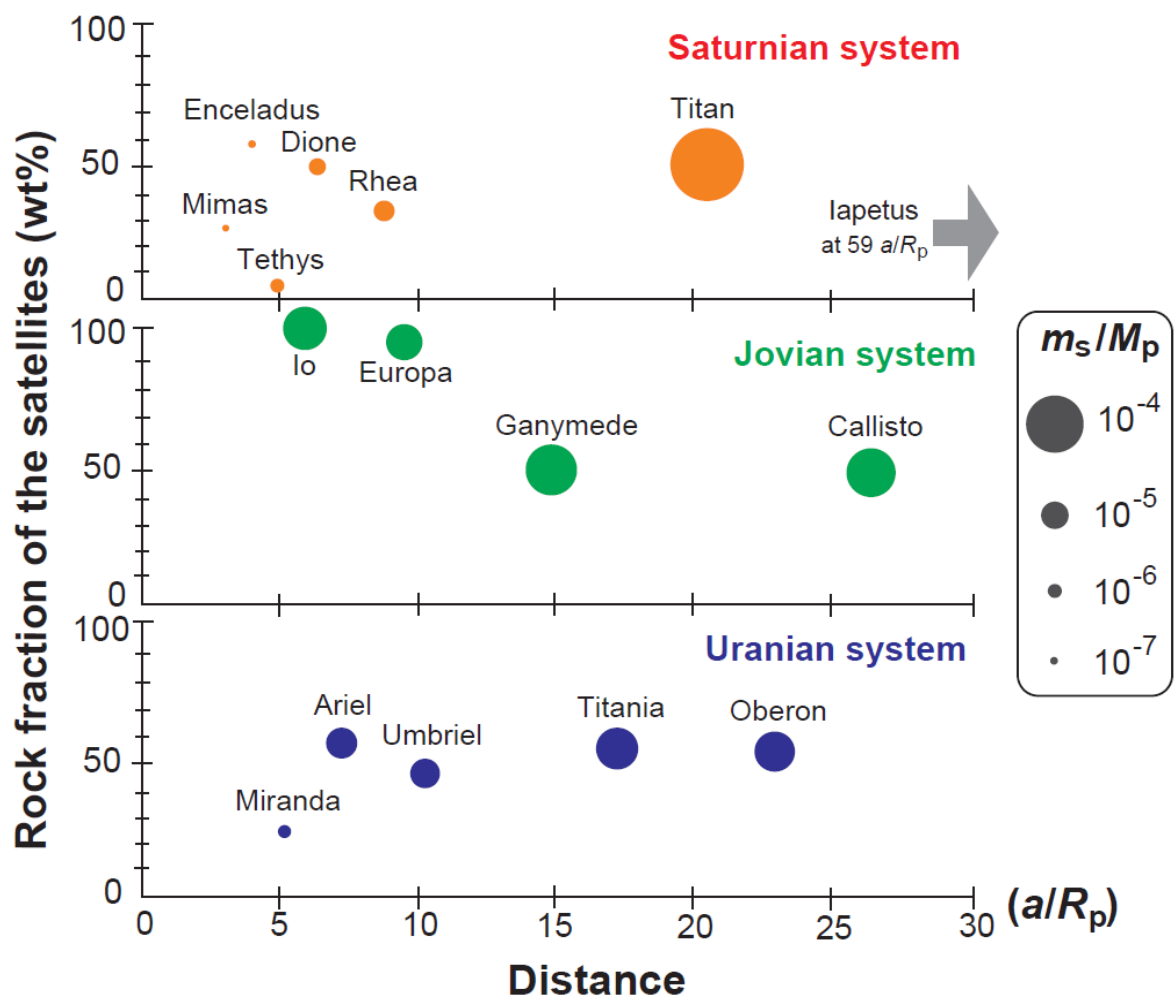

**Figure 1.** A comparison of the satellite systems around the three giant planets for the rock mass fraction, satellite mass, and distance from the planet. The orbital distance from the planet ($a$) is normalized by the radius of the central planet ($R_p$). The size of circle represents the satellite mass ($m_s$) normalized by the central planet mass ($M_p$). Satellite data of the Saturnian system are derived from Matson et al., (2009) and Jovian and Uranian from Beatty (1999).

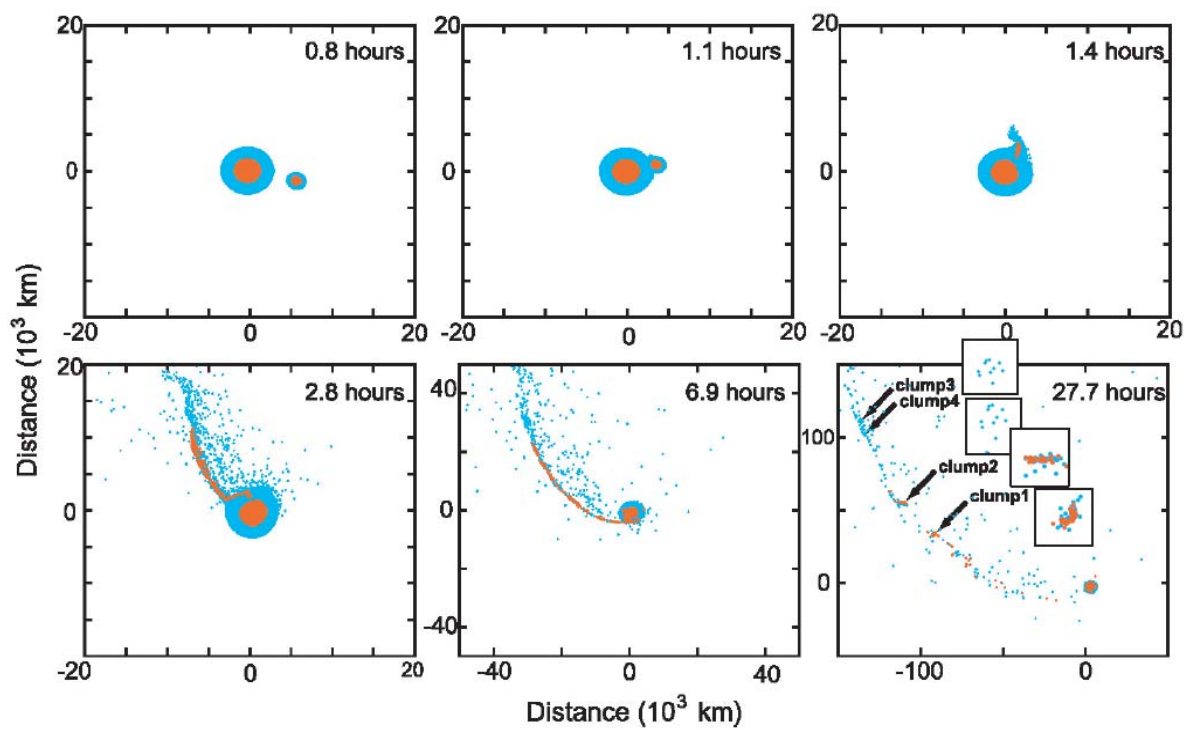

Figure 2 of Sekine & Genda

**Figure 2.** Times series snapshots of a ice-rich and rock-rich satellites forming impact (clump-forming collision). A clump is defined as an aggregate of more than 10 SPH particles, which are gravitationally bounded with each other. Close-up views of clump are also shown in the last panel. Color indicates material type (blue, water ice; orange, rock). Impact velocity and angle are 1.4 times the escape velocity and 45 degree, respectively.

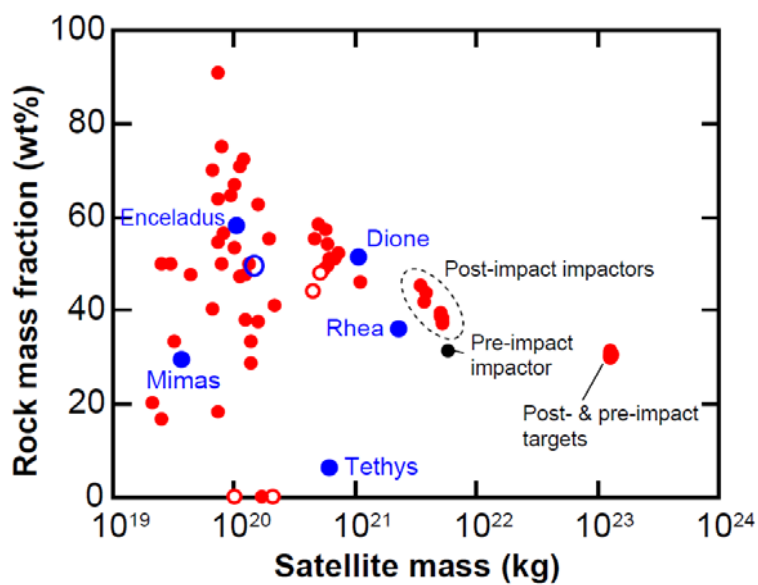

Figure 3 of Sekine & Genda

**Figure 3.** Properties of the Saturnian mid-sized satellites (blue points) compared to the results of our SPH simulations (red points). Red points represent the satellites resulting from giant impacts. Open red points represent the results of the clumps shown in Fig. 2. Open blue point represent the possible rock mass fraction and mass of Enceladus 4.5 Gyrs ago, when the current plume production rate (~200 kg/s: Hansen et al., 2008) has been maintained over the age of the solar system. Black point shows the property of pre-impact impactors. Broken circles represent the properties of the impactor satellites in the aftermath of hit-and-run collisions.